\documentclass[referee]{biom}

\newcommand{\newc}{\newcommand}
\newc{\BE}{ \begin{enumerate} }
\newc{\EE}{ \end{enumerate} }
\newc{\BI}{ \begin{itemize} }
\newc{\EI}{ \end{itemize} }
\newc{\BC}{ \begin{center} }
\newc{\EC}{ \end{center} }
\newc{\bnneq}{\begin{eqnarray*} }
\newc{\enneq}{\end{eqnarray*} }
\newc{\bneq}{\begin{eqnarray} }
\newc{\eneq}{\end{eqnarray} }

\usepackage{epsfig, subfigure,lscape}
\usepackage{amssymb}
\usepackage{epsfig}
\usepackage{rotating}
\usepackage{lscape}
\usepackage{algorithmic}
\usepackage{color}

\def\ci{\perp\!\!\!\perp}

\begin{document}

\title{Learning networks from high dimensional binary data: An
  application to genomic instability data}
\author{Pei Wang, Dennis L. Chao, Li Hsu$^*$ \\
Division of Public Health Sciences, Fred Hutchinson Cancer Research
Center, Seattle, WA\\
$^*$Corresponding Author: 1100 Fairview Ave. N., M2-B500, Fred
Hutchinson Cancer Research \\
Center, Seattle, WA 98109. Email: lih@fhcrc.org. }

\begin{abstract}
Genomic instability, the propensity of aberrations in chromosomes,
plays a critical role in the development of many diseases. High
throughput genotyping experiments have been performed to study genomic instability in diseases. The output of such experiments
can be summarized as high dimensional binary vectors, where each
binary variable records aberration status at one marker locus. It is
of keen interest to understand how these aberrations interact with
each other. In this paper, we propose a novel method,
\texttt{LogitNet}, to infer the interactions among aberration
events. The method is based on penalized logistic regression with an
extension to account for spatial correlation in the genomic
instability data.  We conduct extensive simulation studies and show
that the proposed method performs well in the situations considered.
Finally, we illustrate the method using genomic instability data from
breast cancer samples.
\vspace{0.5in}


\end{abstract}
\maketitle

\section{Introduction}

Genomic instability refers to the propensity of aberrations in
chromosomes such as mutations, deletions and amplifications.  It has
been thought to play a critical role in the development of many
diseases, for example, many types of cancers (Klein and Klein 1985).
Identifying which aberrations
contribute to disease risk, and how they may interact with each other
during  disease development is of keen interest.  High throughput genotyping
experiments have  been performed to interrogate these aberrations
in diseases, providing  a vast amount of information on genomic
instabilities on tens of thousands of marker loci
simultaneously. These data can essentially be organized as a $n \times
p$ matrix where $n$  is the number of samples, $p$ is the
number of marker loci, and the $(i,j)^{th}$ element of the matrix is the
\texttt{binary} aberration status for the $i$th sample at the $j$th
locus. We refer to the interactions among aberrations as oncogenic
pathways. Our goal is to infer oncogenic pathways based on these binary
genomic instability profiles. 

Oncogenic pathways can be compactly represented by graphs, in which
vertices represent $p$ aberrations and edges represent interactions
between aberrations.   Tools developed for graphical models
(Lauritzen 1996) can therefore be employed to infer interactions
among $p$ aberrations.  Specifically, each vertex represents a
binary random variable that codes aberration status at a locus, and
an edge will be drawn between two vertices if the corresponding two
random variables are conditionally dependent given all other random
variables. Here, we want to point out that graphical models based on
conditional dependencies provide information on ``higher order"
interactions compared to other methods (e.g., hierarchical
clustering) which examine the marginal pairwise correlations. The
latter does not tell, for example, whether a non-zero correlation is
due to a direct interaction between two aberration events or due to
an indirect interaction through a third intermediate aberration
event.

There is a rich literature on fitting graphical models for a limited
number of variables (see for example Dawid and Lauritzen 1993;
Whittaker 1990; Edward 2000; Drton and Perlman 2004, and references
therein). However, in genomic instability profiles, the number of
genes $p$ is typically much larger than the number of samples $n$.
Under such high-dimension-low-sample-size scenarios, sparse
regularization becomes indispensable for purposes of both model
tractability and model interpretation. Some work has already been
done to tackle this challenge for high dimensional continuous
variables. For example, Meinshausen and Buhlmann (2006) proposed performing neighborhood selection with \texttt{lasso} regression
(Tibshirani 1996) for each node. Peng et al. (2009a) extended the
approach by imposing the sparsity on the whole network instead of
each neighborhood, and implemented a fast computing algorithm. In
addition, a penalized maximum likelihood approach has been carefully
studied by Yuan and Lin (2007), Friedman et al.(2007b) and Rothman
et al.(2008), where the $p$ variables were assumed to follow a
multivariate normal distribution. Besides these cited works, various
other regularization methods have also been developed for high
dimensional continuous variables (see for example, Li and Gui 2006
and Schafer and Strimmer 2007).  Bayesian approaches have been
proposed for graphical models as well (see for example, Madigan et
al. 1995).

In this paper, we consider binary variables and propose a
novel method, \texttt{LogitNet}, for inferring edges, i.e.,
the conditional dependence between pairs of aberration events given all others.
Assuming a tree topology for oncogenic pathways, we derive
the joint probability distribution of the $p$ binary variables,
which naturally leads to a set of $p$ logistic regression models
with the combined $p\times p$ coefficient matrix being symmetric.
We propose sparse logistic regression with a \texttt{lasso} penalty
term and extend it to account for the spatial correlation
along the genome.   This extension together with the enforcement of
symmetry of the coefficient matrix produces a group selection
effect, which enables \texttt{LogitNet} to make good use of spatial
correlation when inferring the edges.

\texttt{LogitNet} is related to the work by Ravikumar et al.
(2009), which also utilized sparse logistic regression to construct a 
network based on high dimensional binary variables. The basic idea
of Ravikumar et al. is the same as that of Meinshausen and
Buhlmann's (2006) neighborhood selection approach, in which sparse
logistic regression was performed for each binary variable given all
others. Sparsity constraint was then imposed on each neighborhood and
the sparse regression was performed for each binary 
variable separately. Thus, the symmetry of conditional dependence
obtained from regressing variable $X_r$ on variable $X_s$ and from 
regressing $X_s$ on $X_r$ is not guaranteed.  As such, it can yield
contradictory neighborhoods, 
which makes interpretation of the results difficult.  It also loses
power in detecting dependencies, especially when the sample size is
small.  The proposed \texttt{LogitNet}, on the other hand,  makes use
of the symmetry, which produces compatible logistic regression models
for all variables and has thus achieved a more coherent result with
better efficiency than the Ravikumar et al. approach. We show by
intensive simulation studies that \texttt{LogitNet} performs better in
terms of false positive rate and false negative rate of  edge
detection. 


The rest of the paper is organized as follows. In section 2, we will
present the model, its implementation and the selection of the
penalty parameter. Simulation studies of the proposed method and the
comparison with the Ravikumar et al. approach are described in
Section 3. Real genomic instability data from breast cancer samples is used
to illustrate the method and the results are described in Section 4.
Finally, we conclude the paper with  remarks on future work.

\section{Methods}

\subsection{\texttt{LogitNet} Model and Likelihood Function}
Consider a $p \times 1$ vector of binary variables $X^T = (X_1,
\ldots, X_p)$ for which we are interested in inferring the
conditional dependencies.  Here the superscript $T$ is a transpose.
The pattern of conditional dependencies between these binary
variables can be described by an undirected graph ${\cal G} = (V,
E)$, where $V$ is a finite set of vertices, $(1, \ldots, p)$, that
are associated with binary variables $(X_1, \ldots, X_p)$; and $E$
is a set of pairs of vertices such that each pair in $E$ are
conditionally dependent given the rest of binary variables. We assume
that the edge set $E$ doesn't contain cycles,
i.e., no path begins and ends with the same vertex.  For
example, in a set of four vertices, if the edge set includes (1,2),
(2,3), and (3,4), it can't include the edge (1,4) or (1,3) or (2,4),
as it will form a cycle. Under this assumption, the joint
probability distribution $\Pr(X)$ can be represented as a product of
functions of pairs of binary variables.  We formalize this result in
the following proposition:

\textit{ \noindent \textbf{Proposition 1.} Let $V = \{1, \ldots,
p\}$ and $X_{-(r,s)}$ denote the vector of binary variables $X$
excluding $X_r$ and $X_s$ for $r,s \in V$.  Define the edge set
$$E = \{(r, s) | \Pr(X_r, X_s | X_{-(r,s)}) \ne \Pr(X_{r}|X_{-(r,s)}) \Pr(X_{s} | X_{-(r,s)});  r, s \in V, r < s \},$$
and
$|E| = K$. If ${\cal G}$ doesn't contain cycles, then there exist
functions $\{h_k, k= 1, \ldots, K\}$ such that
\[ \Pr(X) = \prod_{k=1}^K h_k(X_{r_k}, X_{s_k}), \]
where $(r_k, s_k) \in E \textit{ for } k = 1, \ldots, K$. }

The proof of Proposition 1 is largely based on the Hammersley and
Clifford theorem (Lauritzen, 1996) and given in Supplementary Appendix A.

Assuming $\Pr(X)$ is strictly positive for all values of $X$, then
the above probability distribution leads to the well known
quadratic exponential model
\begin{eqnarray}
\Pr(X=x) = \Delta^{-1} \exp(x^T \theta + z^T \kappa), \label{eqn1}
\end{eqnarray}
where $z^T = (x_1x_2, x_1x_3, \ldots, x_{p-1}x_p)$, $\theta^T =
(\theta_1, \ldots, \theta_p)$, $\kappa^T = (\kappa_{12},
\kappa_{13}, \ldots, \kappa_{(p-1)p})$, and $\Delta$ is a
normalization constant such that $\Delta = \sum_{x_1 = 0}^1 \cdots
\sum_{x_p=0}^1 \exp(x^T \theta + z^T \kappa)$.

Under this model, the zero values in $\kappa$ are equivalent to the
conditional independence for the corresponding binary variables. The
following proposition describes this result and the proof is given
in Supplementary Appendix B.

\textit{ \noindent \textbf{Proposition 2.} If the distribution on
$X$ is (\ref{eqn1}), then    $X_r \ci X_s  \,|\, X_{-(r,s)}$ if and
only if $\kappa_{rs} = 0$,   for $1\leq r<s \leq p$.}

As the goal of graphical model selection is to infer the edge set $E$
which represents the conditional dependence among all
the variables, the result of Proposition 2 implies that we can infer
the edge between a pair of events, say $X_r$ and $X_s$,
based on whether or not $\kappa_{rs}$ is equal to 0. Interestingly,
under model (\ref{eqn1}), $\kappa$ can also be interpreted as
a conditional odds ratio.  This can be seen from
\begin{eqnarray*}
\lefteqn{\frac{\Pr(x_s = 1 | x_1, \ldots, x_{s-1}, x_{s+1}, \ldots,
  x_p)}{\Pr(x_s = 0 | x_1, \ldots, x_{s-1}, x_{s+1}, \ldots, x_p)}} \\
&  = &
 \exp(\kappa_{1s} x_1 + \ldots + \kappa_{(s-1)s} x_{s-1} + \theta_s
+ \kappa_{s(s+1)} x_{s+1} + \ldots + \kappa_{sp} x_p).
\end{eqnarray*}
Taking the log transformation of the left hand side of this equation
results the familiar form of a logistic regression model, where the
outcome is  the $j$th binary variable and the predictors are all
the other binary variables. Doing this for each of $x_1, x_2, \ldots,
x_p$, we obtain $p$ logistic regressions models:
\begin{eqnarray}\label{eqn:Plogistic}
\left \{ \begin{array}{rcl}
\mbox{logit}\{\Pr(x_1 = 1 | x_2, \ldots,   x_p)\} & = & \theta_1 +
\kappa_{12} x_2 + \ldots + \kappa_{1p} x_p,  \\
& \vdots &  \\
\mbox{logit} \{\Pr(x_p = 1 | x_1, \ldots,   x_{p-1})\} & = &
\kappa_{1p} x_1 + \ldots + \kappa_{(p-1)p} x_{p-1} + \theta_p.
\end{array} \right .
\end{eqnarray}
The matrix of all of the regression coefficients from $p$ logistic
regression models can then be row combined as
\[
\cal{B} =\left (\begin{array}{cccc}
\theta_1 & \kappa_{12} & \ldots & \kappa_{1p} \\
\kappa_{12} & \theta_2 & \ldots & \kappa_{2p} \\
\vdots & &\ddots & \\
\kappa_{1p} & \ldots & \ldots & \theta_p
\end{array} \right )
\]
with matrix elements defined by $\beta_{rs}$ for the $r$th row and
the $s$th column of ${\cal B}$.  It is easy to see that the
$\cal{B}$ matrix is symmetric, i.e.,
$\beta_{rs}=\beta_{sr}=\kappa_{rs}$, $i \ne j$ under model (\ref{eqn:Plogistic}).  Vice
versa, the symmetry of $\cal{B}$ ensures the compatibility of
the $p$ logistic conditional distributions (\ref{eqn:Plogistic}), and the
resulting joint distribution is the quadratic exponential model
(\ref{eqn1})(Joe and Liu, 1986). Thus, to infer the edge set $E$ of
the graphical model, i.e., non-zero off-diagonal entries in
$\cal{B}$, we can resort to regression analysis by simultaneously
fitting the $p$ logistic regression models in~(\ref{eqn:Plogistic})
with symmetric $\cal{B}$.

Specifically, let $X_{n \times p}$ denote the data which consists of
$n$ samples each measured with $p$-variate binary events.  We also
define two other variables mainly for the ease of the presentation
of the likelihood function: (1) $Y$ is the same as $X$ but with 0s replaced
with -1s;  (2) $X^r, r = 1, \ldots, p$ same as X but with $r$th
column set to 1. We propose to maximize the joint log likelihood of
the $p$ logistic regressions in~(\ref{eqn:Plogistic}) as follows:
\begin{eqnarray}\label{eqn:JointLogit}
l ({\cal{B}}) = -\sum_{r=1}^p \sum_{i=1}^n \log \left\{ 1 +
\exp(-X^r[i,]{\cal{B}}[r,]^T \cdot y_{ir})\right\}.
\end{eqnarray}
where $X^r[i,] = (x^r_{i1}, \ldots, x^r_{ip})$; and ${\cal{B}}
[r,]=(\beta_{r1}, ..., \beta_{rp})$. Note, here we have the
constraints $\beta_{rs}=\beta_{sr}$ for $1 \leq r < s \leq p$; and
$\beta_{rr}$ now represents the intercept $\theta_r$ of the $r$th
regression.

Recall that our interest is to infer oncogenic pathways based
on genome instability profiles of tumor samples. Most often, we are dealing with
hundreds or thousands of genes and only tens or hundreds of samples.
Thus, regularization on parameter estimation becomes indispensable
as the number of variables is larger than the sample size, $p>n$. In
the past decade, $\ell_1$ norm based sparsity constraints such as
\texttt{lasso} (Tibshirani 1996) have shown considerable success in
handling high-dimension-low-sample-size
problems when the true model is sparse relative to the
dimensionality of the data. Since it is widely believed that genetic
regulatory relationships are intrinsically sparse (Jeong et al.
2001; Gardner et al. 2003), we propose to use $\ell_1$ norm penalty
for inferring oncogenic pathways.  The penalized loss
function can be written as:
\begin{equation}\label{eqn:PenalizedJL}
l_{\lambda}^{lasso}({\cal{B}}) = -l({\cal{B}}) + \lambda \sum_{r=1}^p
\sum_{s=r+1}^p |\beta_{rs}|.
\end{equation}
Note that $\ell_1$-norm penalty is imposed on all off-diagonal entries of
$\cal{B}$ matrix simultaneously to control the overall sparsity
of the joint logistical regression model, i.e., only a limited number
of $\beta_{rs}$, $r \ne s$  will be non-zero. We then estimate
$\cal{B}$ by $\widehat{\cal{B}}(\lambda):= \textrm{arg min}_{\cal{B}}
l_{\lambda}^{lasso}({\cal{B}})$. In the rest of the paper, we refer to the
model defined in~(\ref{eqn:PenalizedJL}) as \texttt{LogitNet} model,
$\widehat{\cal{B}}(\lambda)$ as the \texttt{LogitNet}
estimator and $\widehat\beta_{rs}(\lambda)$ as the $rs$th element of
$\widehat{\cal{B}}(\lambda)$.

As described in the Introduction, the \texttt{LogitNet} model is
closely related to the work by Ravikumar et al. (2009) which fits
$p$ \texttt{lasso} logistic regressions separately (hereafter
referred to as \texttt{SepLogit}).  Our model, however, differs in
two aspects: (1) \texttt{LogitNet} imposes the \texttt{lasso}
constraint for the entire network while \texttt{SepLogit} does it
for each neighborhood; (2) \texttt{LogitNet} enforces symmetry when
estimating the regression coefficients while \texttt{SepLogit}
doesn't,  so for \texttt{LogitNet} there are only about half of the
parameters needed to be estimated as for \texttt{SepLogit}. As a
result, the \texttt{LogitNet} estimates are more efficient and the
results are more interpretable than \texttt{SepLogit}.


\subsection{Model fitting}
In this section, we describe an algorithm for obtaining the
\texttt{LogitNet} estimator $\widehat{\cal{B}}(\lambda)$.  The
algorithm extends the gradient descent  algorithm (Genkin et al. 2007)
to enforce the symmetry of $\cal{B}$. 
Parameters are  updated one at a time using a one-step Newton-Raphson
algorithm, in the same spirit as the shooting algorithm (Fu, 1998) and the
coordinate descent algorithm (Friedman et al., 2007a) for
solving the general linear \texttt{lasso} regressions.

More specifically, let $\dot l(\beta_{rs})$ and $\ddot l(\beta_{rs})$
be the first- and second- partial derivatives of log-likelihood
$l(\cal{B})$ with respect to $\beta_{rs}$,
\begin{eqnarray*}
\dot l(\beta_{rs}) & = & \sum_{i=1}^n\frac{X^r[i,s]
  Y[i,r]}{1+\exp(R_{r})} +
  \sum_{i=1}^n\frac{X^s[i,r]Y[i,s]}{1+\exp(R_{s})}, \\
\ddot l(\beta_{rs}) & = & \sum_{i=1}^n(X^r[i,s])^2
\frac{\exp(R_{r})}{\{1+\exp(R_{r})\}^2} +
  \sum_{i=1}^n  (X^s[i,r])^2 \frac{\exp(R_{s})}{\{1+\exp(R_{s})\}^2},
\end{eqnarray*}
where $R_{r} = X^r[i,] \beta^T[r,]Y[i,r]$.  Under the
Newton-Raphson algorithm, the update for the estimate $\widehat
\beta_{rs}$ is $\Delta
\beta_{rs}=-\dot l (\beta_{rs})/\ddot l(\beta_{rs})$.   For the
penalized likelihood (\ref{eqn:PenalizedJL}), the update for
$\widehat \beta_{rs}(\lambda)$ is
\begin{eqnarray}
 \Delta \beta_{rs}^{lasso} & = & -\frac{\dot l_\lambda^{lasso}(\beta_{rs})}{\ddot
l_\lambda^{lasso}(\beta_{rs})}  \nonumber \\
& = & \Delta \beta_{rs} - \frac{\lambda}{\ddot l(\beta_{rs})} \mbox{sgn}(\beta_{rs}),
\label{update}
\end{eqnarray}
where $\mbox{sgn}(\beta_{rs})$ is a sign function, which is 1 if
$\beta_{rs}$ is positive and  -1 if $\beta_{rs}$ is negative. The
estimates are also thresholded such that if an update overshoots and
crosses the zero, the update will be set to 0.  If the current
estimate is 0, the algorithm will try both directions by setting
$\mbox{sgn}$ to be 1 and -1, respectively.  By the convexity of
(\ref{eqn:PenalizedJL}), the update for both directions can not be
simultaneously successful. If it fails on both directions, the estimate will
be set to 0. The algorithm also takes other steps to make sure
the estimates and the numerical procedure are stable, including
limiting the update sizes and setting the upper bounds for $\ddot l$
(Zhang and Oles 2001).  See Supplemental Appendix C for more
details of the algorithm.


To further improve the convergence speed of the algorithm, we
utilize the \texttt{Active-Shooting} idea proposed by Peng et al.
(2009a) and Friedman et al. (2009).  Specifically, at each
iteration, we define the set of currently non-zero
coefficients 
as the current active set and conduct the following two steps: (1)
update the coefficients in the active set until convergence is
achieved;  (2) conduct a full loop update on all the coefficients
one by one.  We then repeat (1) and (2) until convergence is
achieved on all of the coefficients. Since the target model in our
problem is usually very sparse,  this algorithm achieves a very
fast convergence rate by focusing on the small subspace whose
members are more likely to be in the model.

We note that in equation (5) the regularization shrinks the estimate
towards zero by the amount determined by the penalty parameter
$\lambda$ and that each parameter is not penalized by the same
amount: $\lambda$ is weighted by the variance ${\ddot
  l(\beta_{rs})}^{-1}$ of $\widehat \beta_{rs}$.  In other words,
parameter estimates that have larger variances will be penalized
more than the ones that have smaller variances.  It turns out that
this type of penalization is very useful, as it would also offer us
ways to account for the other features of the data. In the next
section we show a proposal for adding another weight function to
account for spatial correlations in genomic instability data.

\subsection{Spatial correlation}
Spatial correlation of aberrations is common in genomic
instability data.  When we perform the regression of $X_r$ on all
other binary variables, loci that are  spatially closest to
the $X_r$ are likely the strongest predictors in the model and will
explain away most of the variation in $X_r$.  The loci at the other
locations of the same or other chromosomes, even if they are
correlated with $X_r$,  may be left out in the model.  Obviously
this result is not desirable because our objective is to identify the
network among all of these loci (binary variables), in particular
those that are not close spatially as we know them already.

One approach to accounting for this undesirable spatial
effect is to downweight the effect of the neighboring loci of $X_r$
when
regressing $X_r$ on the rest of the loci.  Recall that in Section
2.2, we observed that the penalty term in (\ref{update}) is
inversely weighted by the variance of the parameter estimates.
Following the same idea, we can achieve the downweighting of neighboring
loci by letting the penalty term be proportional to the strength of
their correlations with $X_r$.   This way we can
shrink the effects of the neighboring loci with strong
spatial correlation more than those that have less or no spatial
correlation.  Specifically, the update for the parameter estimate
$\beta_{rs}$ in (\ref{update}) can be written as
\begin{eqnarray*}
 \Delta \beta_{rs}^{lasso} & = &
\Delta \beta_{rs} - \lambda \frac{w_{rs}}{\ddot l(\beta_{rs})}
\mbox{sgn}(\beta_{rs}),
\end{eqnarray*}
where $w_{rs}$ is the weight for the spatial correlation.  Naturally
the weight $w_{rs}$ for $X_r$ and $X_s$ on different
chromosomes is 1 and for $X_r$ and $X_s$ on
the same chromosome should depend on the strength of the spatial
correlation.  As the spatial correlation varies across the genome, we
propose the following adaptive estimator for
$w_{rs}$:



\begin{enumerate}
\item Calculate the odds ratio $\alpha$ between every locus in the
chromosome with the target locus by univariate logistic regression.
\item Plot the $\alpha$'s by their genomic locations and smooth the
  profile by loess with a window size of 10 loci.
\item Set the smoothed curve $\widetilde{\alpha}$ to 0 as soon as the
curve starting from the target locus hits 0. Here ``hits 0" is
defined as $\widetilde{\alpha}<\varepsilon$, where
$\varepsilon=\textrm{median}_r|\widetilde{\alpha}_r-\widetilde{\alpha}_{r+1}|$.
\item Set the weight $w = \exp(\alpha)$.
\end{enumerate}


It is worth noting that the above weighting scheme together
with the enforcement of the symmetry  of $\cal{B}$ in
\texttt{LogitNet} encourages a group selection effect, \textit{i.e.},
highly correlated predictors tend to be in or out of the model
simultaneously. We illustrate this point with a simple example system
of three variables $X_1, X_2$, and  $X_3$.  Suppose that $X_2$ and
$X_3$ are very close on the genome and highly correlated; and
$X_1$ is associated with $X_2$ and $X_3$ but sits on a different
chromosome.  Under our proposal, the weight matrix $w$ is 1 for all
entries except $w_{23}=w_{32}=a$, which is a large value because of
the strong  spatial correlation between $X_2$ and $X_3$.
Then, for \texttt{LogitNet}, the joint logistic regression model
\begin{eqnarray}
\label{eqn:WeiExa:1}\texttt{logit}(X_1)\sim \beta_{11} +\beta_{12}
X_2 + \beta_{13} X_3,\\
\label{eqn:WeiExa:2} \texttt{logit}(X_2)\sim \beta_{12} X_1
+\beta_{22} + \beta_{23}
X_3,\\
\label{eqn:WeiExa:3}\texttt{logit}(X_3)\sim \beta_{13} X_1
+\beta_{23}X_2 + \beta_{33},
\end{eqnarray}
is subject to the constraint $|\beta_{12}|+|\beta_{13}|+
a|\beta_{23}|<s$.   Because of the large value of $a$, $\beta_{23}$
will likely be shrunk to zero,  which ensures  $\beta_{12}$ and
$\beta_{13}$ to be nonzero in (\ref{eqn:WeiExa:2}) and
(\ref{eqn:WeiExa:3}), respectively.  With the symmetry constraint
imposed on $\cal{B}$ matrix, we also enforce both $\beta_{12}$ and
$\beta_{13}$ to be selected in (\ref{eqn:WeiExa:1}).   This grouping
effect would not happen if we fit only the model
(\ref{eqn:WeiExa:1}) for which only one of $\beta_{12}$ and
$\beta_{13}$ would likely be selected (Zou and Hastie 2005), nor would it
happen if we didn't have a large value of $a$ because  $\beta_{23}$ would
have been the dominant coefficient in models (\ref{eqn:WeiExa:2}) and
(\ref{eqn:WeiExa:3}).   Indeed, the group selection effect of
\texttt{LogitNet} is clearly observed in the simulation studies conducted
in Section 3. 

\subsection{Penalty Parameter Selection}
We consider two procedures for selecting the penalty parameter
$\lambda$: cross validation (CV) and Bayesian Information Criterion
(BIC).
\vspace{-10pt}
\subsubsection{Cross Validation}
After we derive the weight matrix $w$ based on the whole data set,
we divide the data into $V$ non-overlapping equal subsets. Treat the
$v^{th}$ subset $X^{(v)}$ as the $v^{th}$ test set, and its
complement $X^{-(v)}$ as the $v^{th}$ training set. For  a given
$\lambda$, we first obtain the \texttt{LogitNet}
estimate $\widehat{\cal{B}}^{v}(\lambda)$ with the weight matrix $w$
on the $v^{th}$ training set $X^{-(v)}$. Since in our problem the
true model is usually very sparse, the degree of regularization
needed is often high.  As a result, the value of
$\widehat{\cal{B}}^{v}(\lambda)$ could be shrunk far from the true
parameter values.  Using such heavily shrunk estimates for choosing
$\lambda$ from cross validation often results in severe over-fitting
(Efron et al. 2004). Thus, we re-estimate $\cal{B}$ using the
selected model in the $v$th training set without any shrinkage and
use it in calculating the log-likelihood for the $v^{th}$ test set.
The \textit{un-shrunk estimates}
$\widehat{\cal{B}}_{uns}^{(v)}(\lambda)$ can be  easily obtained
from our current algorithm  for the regularized estimates 
with modifications described below:
\begin{enumerate}
\item Define a new weight matrix $\widetilde{w}^{v}$ such that
  $\widetilde{w}_{rs}^{v}=1$, if $\widehat{\beta}_{rs}^{(v)}\neq 0$;
  and $\widetilde{w}_{rs}^{v}=M$, if $\widehat{\beta}_{rs}^{(v)}= 0$,
where $M=\textrm{max}\{w_{rs}\}$.
\item Fit the \texttt{LogitNet} model using the new weight matrix
$\tilde{w}$, thus $\{\beta_{rs}| \widetilde{w}_{rs}^{v}=1\}$ are not
penalized in the model and all other $\beta_{rs}$ are shrunk to 0.
The result is $\widehat{\cal{B}}_{uns}^{(v)}(\lambda)$.
\end{enumerate}
We then calculate the joint log likelihood of logistic regressions
using the un-shrunk estimates on the $v^{th}$ test set
$l(\widehat{\cal{B}}_{uns}^{(v)}(\lambda)|X^{(v)})$ according to
formula (\ref{eqn:Plogistic}).  The optimal
$\lambda_{\textrm{cv}}=\textrm{arg}\max_{\lambda} \sum_{v=1}^V
l(\widehat{\cal{B}}_{uns}^{(v)}(\lambda)|X^{(v)})$.

In order to further control the false positive findings due to
stochastic variation, we employ the \texttt{cv.vote} procedure
proposed by Peng et al. (2009b). The idea is to derive the
``consensus" result of the models estimated from each training set,
as variables that are consistently selected by different training
sets should be more likely to  appear in the true model than the
ones that are selected by one or few training sets.  Specifically,
for a pair of $r$th and $s$th variables, we define
\begin{equation}
s_{rs}(\lambda_{\textrm{cv}})=\left\{ \begin{array}{ll} 1, & \textrm{if }
\sum_{v=1}^{V} I(\widehat{\beta}_{uns,rs}^{(v)}(\lambda_{\textrm{cv}})\neq 0) > V/2;\\
0, & \textrm{otherwise.}
\end{array}
 \right.
\end{equation}
We return $\{s_{rs}(\lambda_{\textrm{cv}})\}$ as our final result.

\subsubsection{BIC}
We can also use BIC to select $\lambda$:
\begin{equation}\label{eqn:BIC}
\lambda_{\textrm{BIC}}=\arg
\min_{\lambda}\left\{-2l(\widehat{\cal{B}}_{uns}(\lambda)|X)+
\textrm{log}(n) \sum_{r<s} I
(\widehat{\beta}_{uns,rs}^{(v)}(\lambda)\neq 0)\right\}
\end{equation}
where $\sum_{r<s} I (\widehat{\beta}_{uns,rs}^{(v)}(\lambda)\neq 0)$
gives the dimension of the parameter space of the selected model.
Here again, \textit{un-shrunk estimates}
$\widehat{\cal{B}}_{uns}^{(v)}(\lambda)$ is used to calculate the log
likelihood.

\section{Simulation Studies}
In this section, we investigate the performance of the
\texttt{LogitNet} method and compare it with \texttt{SepLogit}
which fits $p$ separate lasso logistic regressions all using the same
penalty parameter value (Ravikumar et al., 2009). We use R package
\textit{glmnet} to compute the \texttt{SepLogit} solution and the
same weight matrix as described in Section 2.3 to account for the
spatial correlation. In addition, since the \texttt{SepLogit} method
does not ensure the symmetry of the estimated $\mathcal{B}$ matrix,
there will be cases that $\beta_{rs} =0 $ but $\beta_{sr} \neq 0$ or
vice versa.  In these cases we interpret the result using the ``OR"
rule: $X_r$ and $X_s$ are deemed to be conditionally dependent if
either $\beta_{rs}$ or $\beta_{sr}$ is 0.  We have also used the
``AND" rule, \textit{i.e.} $X_r$ and $X_s$ are deemed to be
conditionally dependent if both $\beta_{rs}$ and
$\beta_{sr}$ are 0.  The ``AND" rule always yields very high false
negative rate. Due to space limitations, we omit the results for the
``AND'' rule.

\subsection{Simulation setting}
We generated background aberration events with spatial correlation
using a homogenous Bernoulli Markov model.  It is part of the
instability-selection model (Newton et al. 1998), which hypothesizes
that the genetic structure of a progenitor cell is subject to
chromosomal instability that causes random aberrations. The Markov
model has two parameters: $\delta$ and $\nu$, where $\delta$ is the
marginal (stationary) probability at a marker locus and $\nu$
measures the strength of the dependence between the aberrations. So
$\delta$ plays the role of a background or sporadic aberration when
$\nu$ affects the overall rate of change in the stochastic process.
Under this model, the infinitesimal rate of change from no
aberration to aberration is $\nu \delta$, and from aberration to no
aberration is $\nu(1-\delta)$. We then super-imposed the aberrations
at disease loci, which were generated according to a pre-determined
oncogenic pathway, on the background aberration events.  The
algorithm for generating an aberration indicator vector $X^T=(X_1,
\cdots, X_p)$ is given below: \vspace{-10pt}



\begin{enumerate}
\item[1.] Specify the topology of an oncogenic pathway for the
  disease loci and the transitional probabilities among the aberrations on the
  pathway. The $K$ disease loci are indexed by $\{{s_1},  \cdots,
  {s_K}\}$, where $s_i \in \{1, \ldots, p\}$ for $i= 1, \ldots, K$.

\item[2.] Generate background aberrations denoted by a $p\times 1$
  vector $Z$ according to the homogenous Bernoulli Markov process with
  preselected values of $\delta=0.05$ and $\nu=15$.

\item[3.] Generate aberration events at disease loci following the
  oncogenic pathway specified in Step 1.  This is denoted by a $p
  \times 1$ vector $U$, where indices $\{{s_1},  \cdots,   {s_K}\}$
  are disease loci. If disease locus $s_i$ has an aberration
  ($U_{s_i}=1$), we also 
assign aberrations to its neighboring loci $U_t=1$, for $t \in
[s_i-a_i, s_i+b_i]$, where $a_i$ and $b_i$ are independently sampled
from Uniform$[0, 30]$. The rest of the elements in $U$ are 0.

\item[4.] Combine the aberration events at disease loci and the background
aberrations by assigning $X_i=1 \textrm{ if }U_i+Z_i>0$ and $0 \textrm{ if
}U_i=Z_i=0$, for $i=1, \ldots, p$.
\end{enumerate}

\begin{figure}[h]
\begin{center}
 \subfigure[The Chain Model]{\label{Fig:ChainModel}
\includegraphics[width=15cm, angle=0]{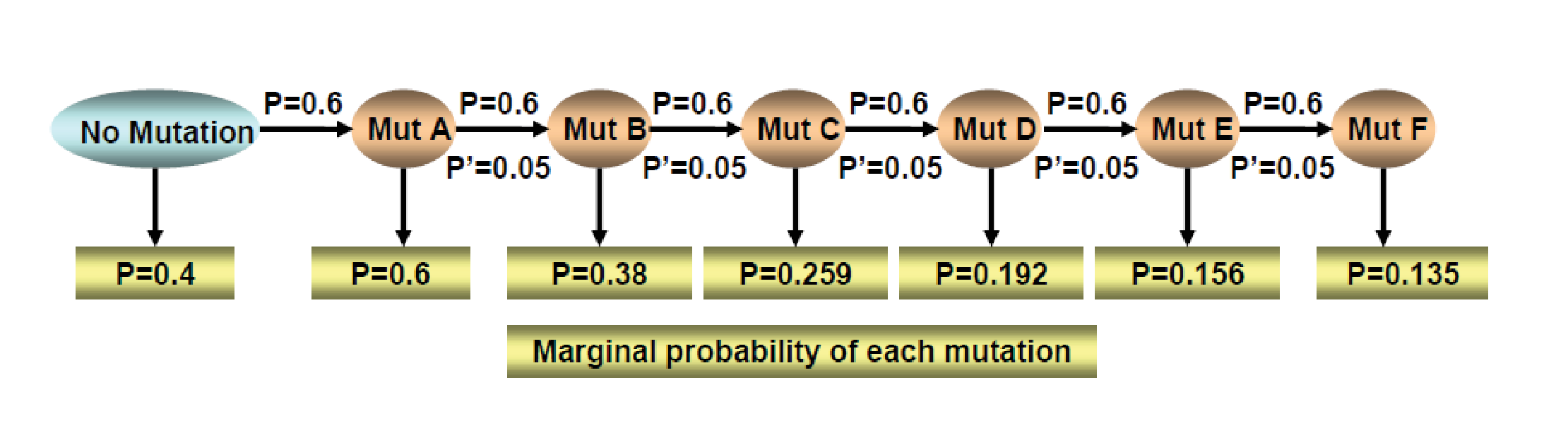}}
\hspace{.0in}

\subfigure[The Tree Model]{ \label{Fig:TreeModel}
 \includegraphics[width=14cm, angle=0]{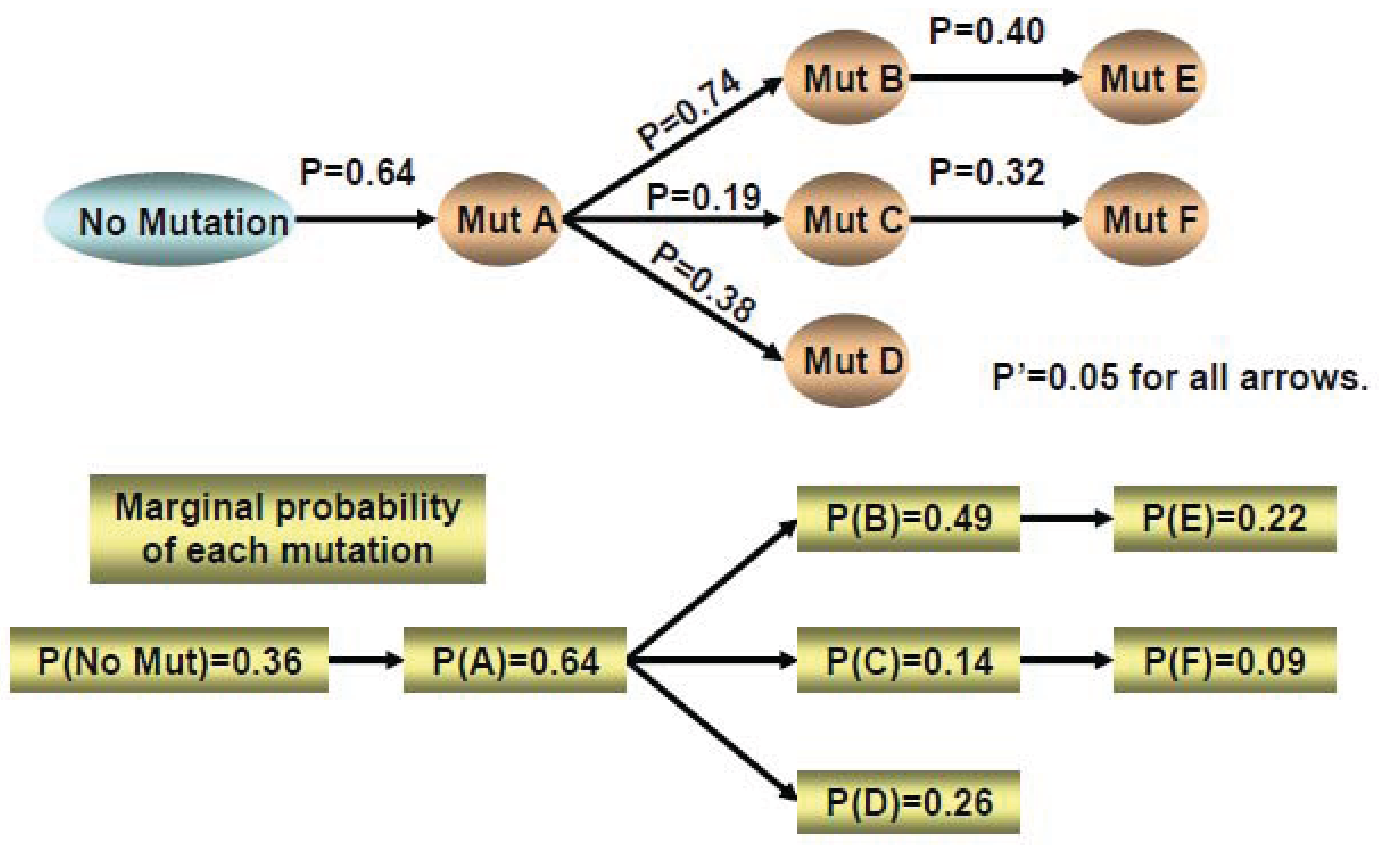}}
\hspace{.0in}
\caption{\textbf{(a)} A chain shape oncogenic pathway;
\textbf{(b)} A tree shape oncogenic pathway. Numbers in the
yellow boxes are the marginal frequencies of the aberration in
the disease population. The numbers on/blow the arrows are the
conditional probabilities between mutations along the oncogenic
pathway. For example, consider the arrow from \texttt{Mut A} to
\texttt{Mut B} in panel \textbf{(a)}, the ``$P=0.6$" above the arrow
means: \texttt{Prob}(Mutation B happens $|$ Mutation A
happens)$=0.6;$ while the ``$P'=0.05$" below the arrow means:
\texttt{Prob}(Mutation B happens $|$ Mutation A does not
happen)$=0.05$. The remark ``$P'=0.05$ for all arrows" in panel
\textbf{(b)} suggests: \texttt{Prob}(The right side mutation happens
$|$ the left side mutation does not happen)$=0.05.$ } \label{Fig1}
\end{center}
\end{figure}

We set $n=200$ and $p=600$ to mimic the dimension of the real data
set used in Section~\ref{sec:RealApplication}, so $V = \{1, \ldots,
600\}$. We assume the $600$ marker loci fall into 6 different
chromosomes with 100 loci on each chromosome.  We consider two
different oncogenic pathway models: a chain shape and a tree shape
(see Figure~\ref{Fig1}). Each model contains 6  aberration events:
$\texttt{M}=\{\texttt{A}, \texttt{B}, \texttt{C}, \texttt{D},
\texttt{E}, \texttt{F}\}$. Without loss of generality, we assume
these 6 aberrations are located in the middle of each chromosome, so
the indices of \texttt{A}--\texttt{F} are $s_{\mathtt{A}} = 50$,
$s_{\mathtt{B}}=150$, $\cdots$, $s_\mathtt{F} = 550$, respectively.
For any $\texttt{u}\in \texttt{M}$, $X_{s_{\mathtt{u}}}=1$ means
aberration \texttt{u} occurs in the sample. \vspace{15pt}

We evaluate the performance of the methods by two metrics:
the false positive rate (\texttt{FPR}) and the false negative rate
(\texttt{FNR}) of edge detection.  Denote the true edge
set $E =\{(u,v) | X_{u} \mbox{ and } X_v \mbox{ are
  conditionally dependent }, u \in V, v \in V\}$.
We define a non-zero $\hat\beta_{rs}$ a false detection if its genome
location indices $(r,s)$ is far from the indices of any true edge:
\[
|r-I_\texttt{u}|+|s-I_\texttt{v}|> 30, \qquad
\forall(\texttt{u},\texttt{v})\in E.
\]
For example, in
Figure~\ref{Fig:ChainBeta} red dots that do not fall into any grey diamond  are
considered false detection. A cutoff 
value of 30 is used here because in the simulation setup (see Step 3)
we set the maximum aberration size around the
disease locus to be 30.  Similarly, we define a conditionally dependent pair
$(\texttt{u},\texttt{v})\in E$ is missed, if there is no non-zero
$\beta$ falling in the grey diamond.  We then calculate \texttt{FPR}
as the number of false detections divided by the total number of non-zero
$\hat\beta_{rs}$, $r<s$; and calculate \texttt{FNR} as the number
of missed $(\texttt{u},\texttt{v})\in E$ divided by the size
of $E$.

\subsection{Simulation I --- Chain Model}
\begin{figure}[h]
\begin{center}
 \subfigure[False positive rate (\texttt{FPR}).]{
 \includegraphics[height=14cm, angle=270]{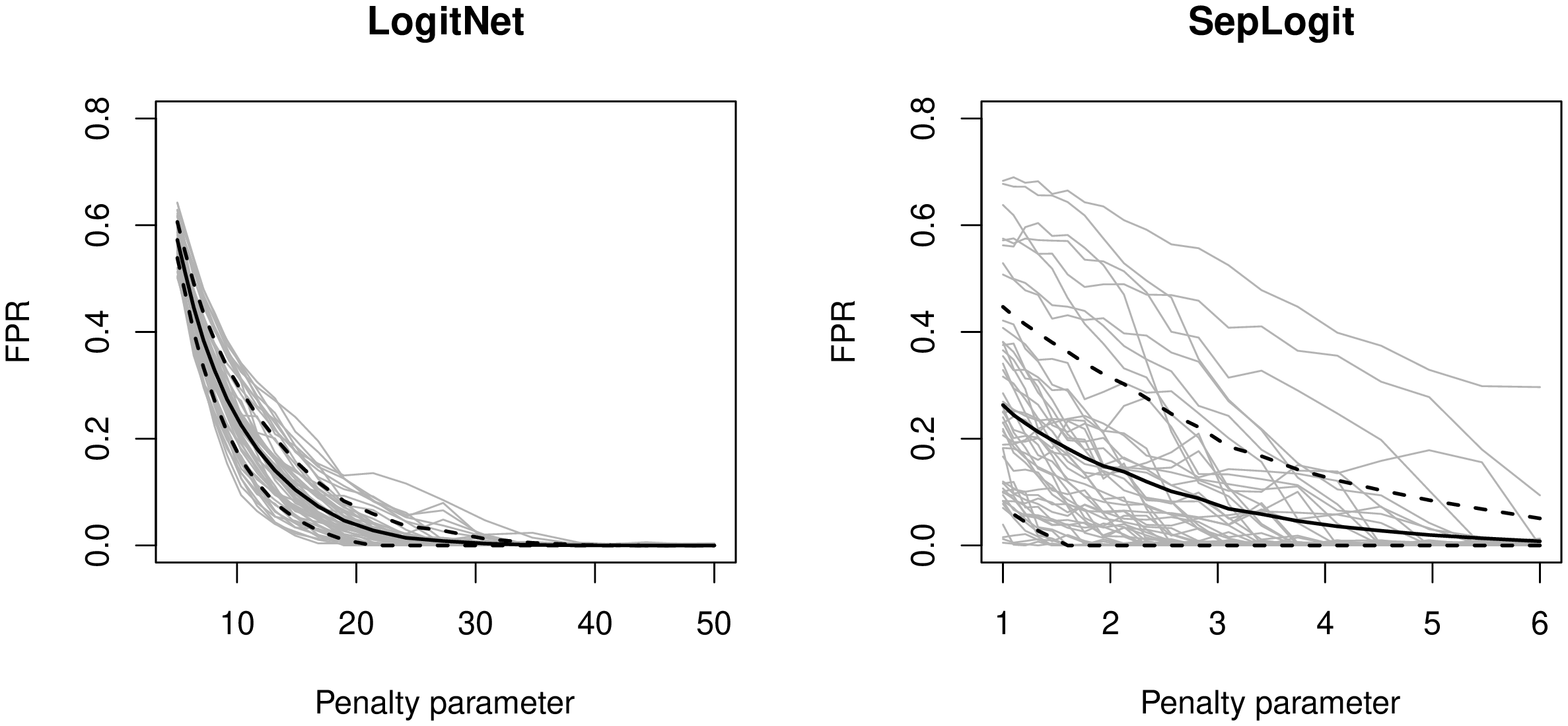}} \hspace{.0in}
\subfigure[False negative rate (\texttt{FNR}).]{\includegraphics[height=14cm,
 angle=270]{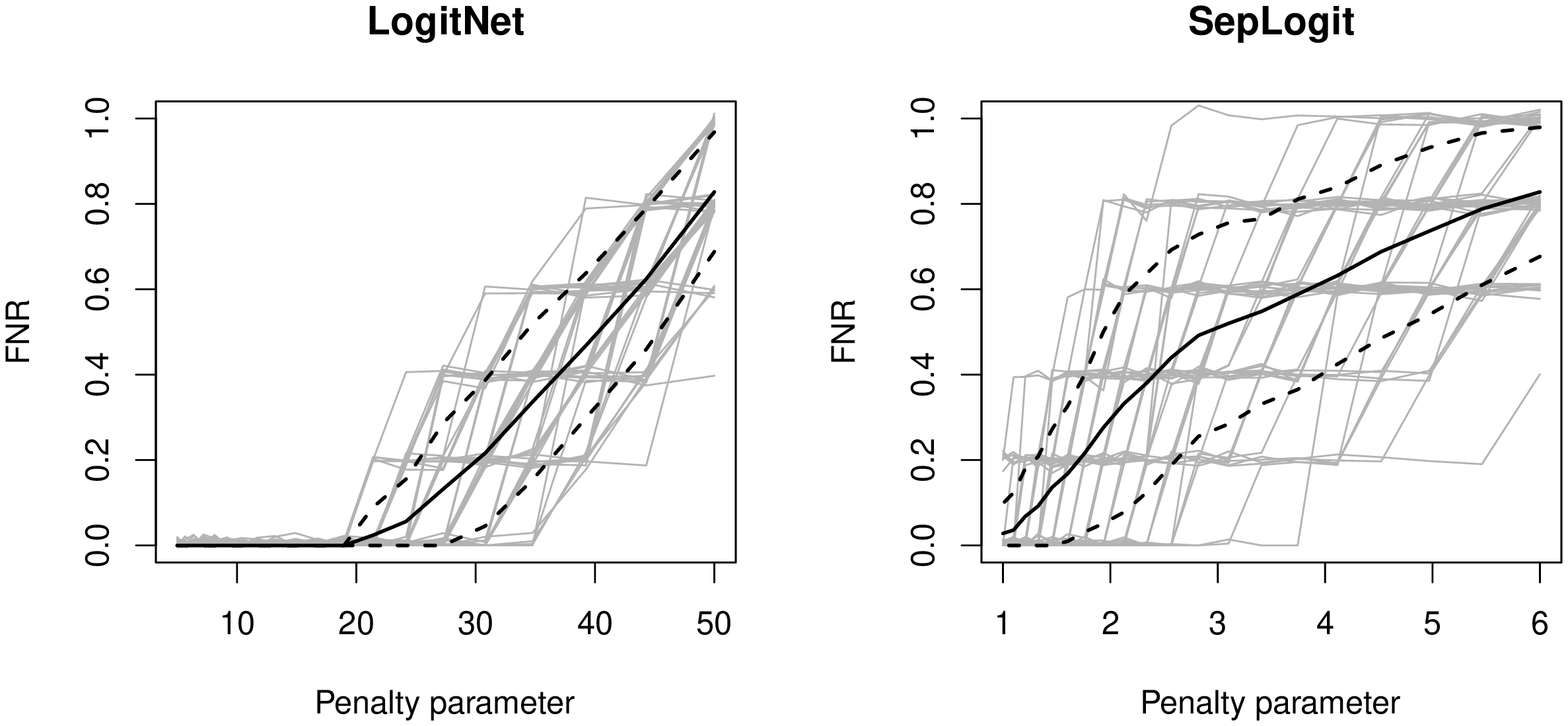}} \hspace{.0in}
 \subfigure[Total error rate (\texttt{FPR}+\texttt{FNR}).]{
 \includegraphics[height=14cm, angle=270]{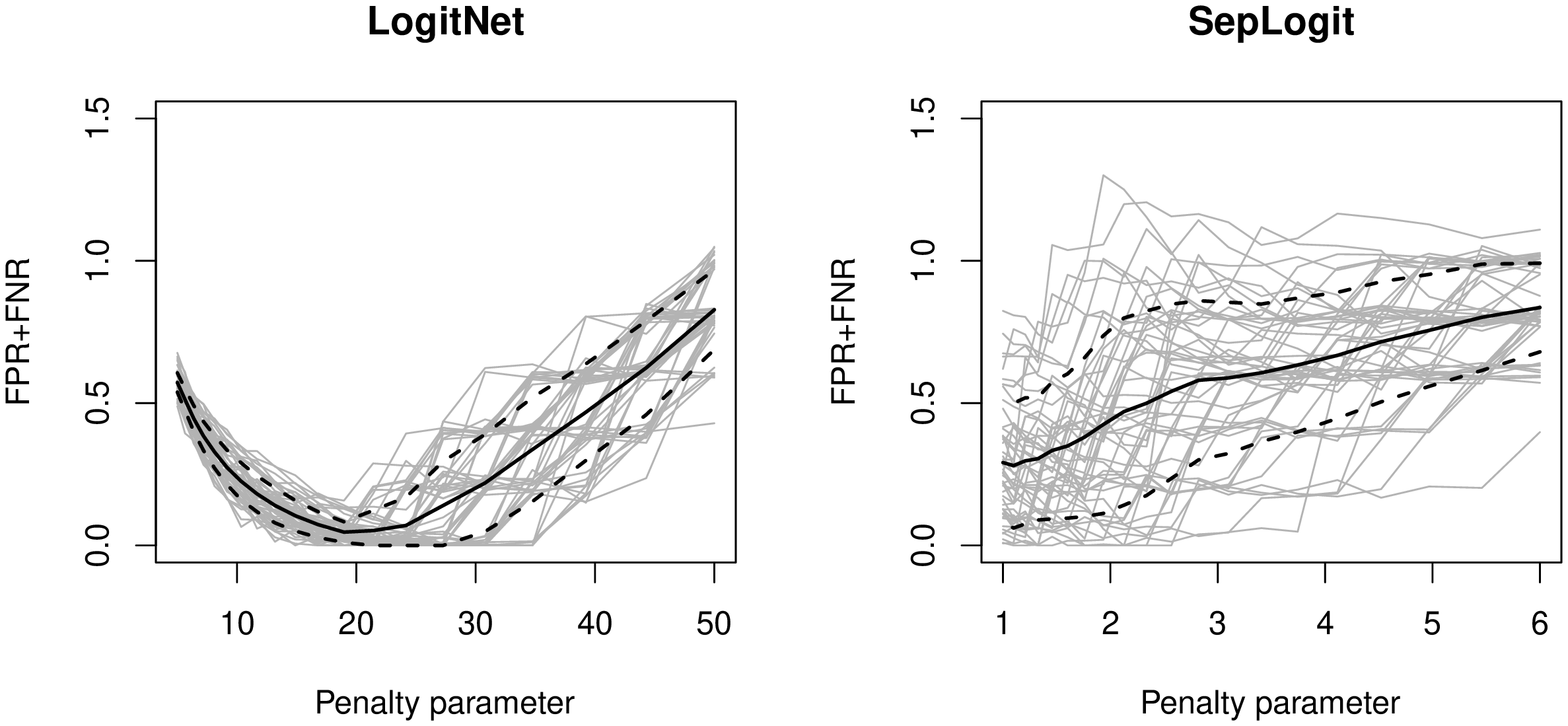}} \hspace{.0in}
\caption{Results of \texttt{LogitNet} and \texttt{SepLogit} for 
the chain model in Figure~\ref{Fig:ChainModel}. Each
grey line represents one of the 50 independent data
sets. The solid line is the mean curve with the two dashed 
lines represent mean $\pm$ one s.d.}
\label{Fig:Chain.error}
\end{center}
\end{figure}

\begin{figure}[h]
\begin{center}
 \subfigure[Estimated $\mathcal{B}$ by \texttt{LogitNet} at the
 $\lambda$ giving the smallest total error rate.]{
 \includegraphics[width=10cm, angle=270]{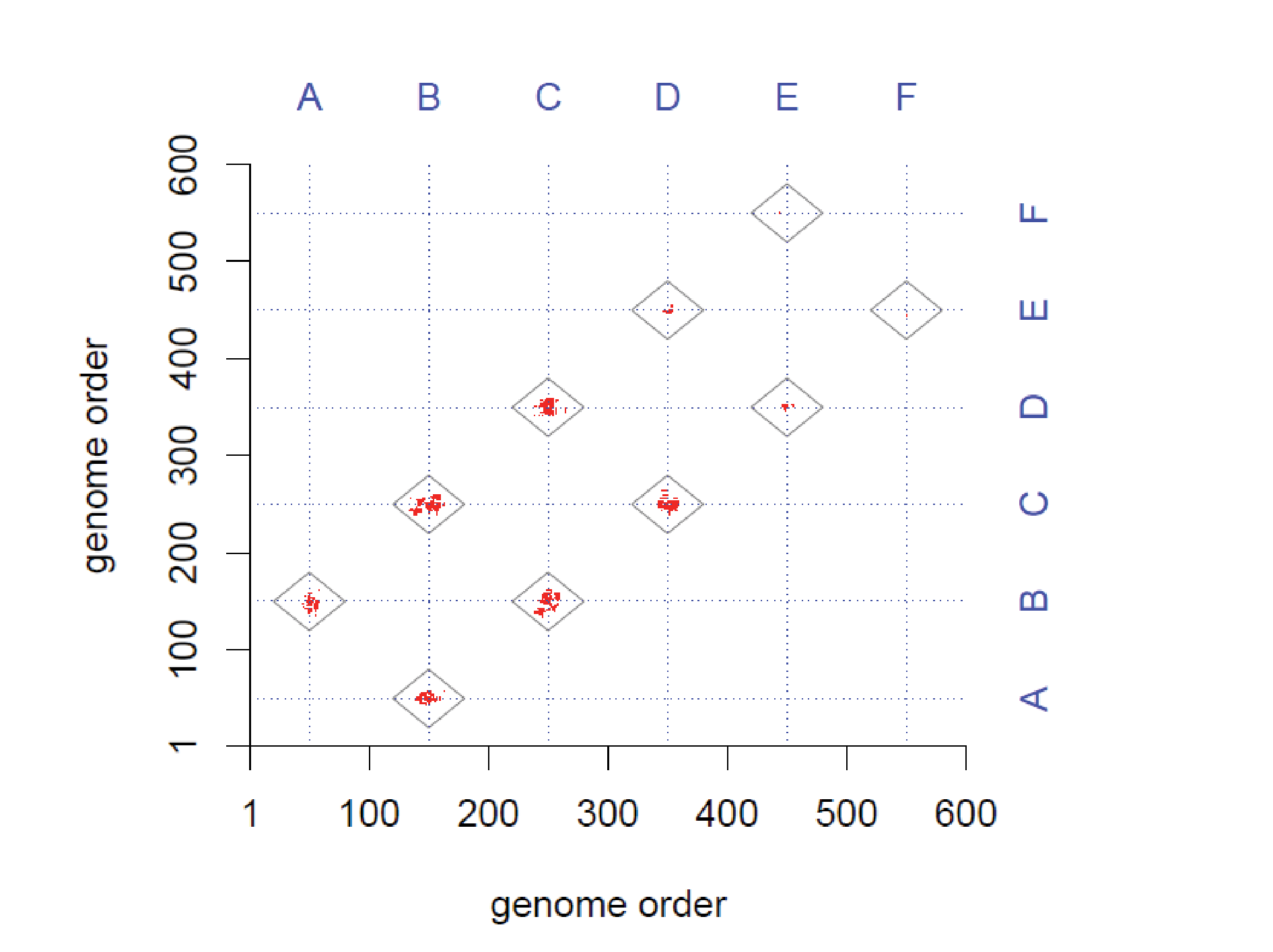}}
\subfigure[Estimated $\mathcal{B}$ by \texttt{SepLogit} at the
$\lambda$ giving the smallest total error rate.]{ 
 \includegraphics[width=10cm, angle=270]{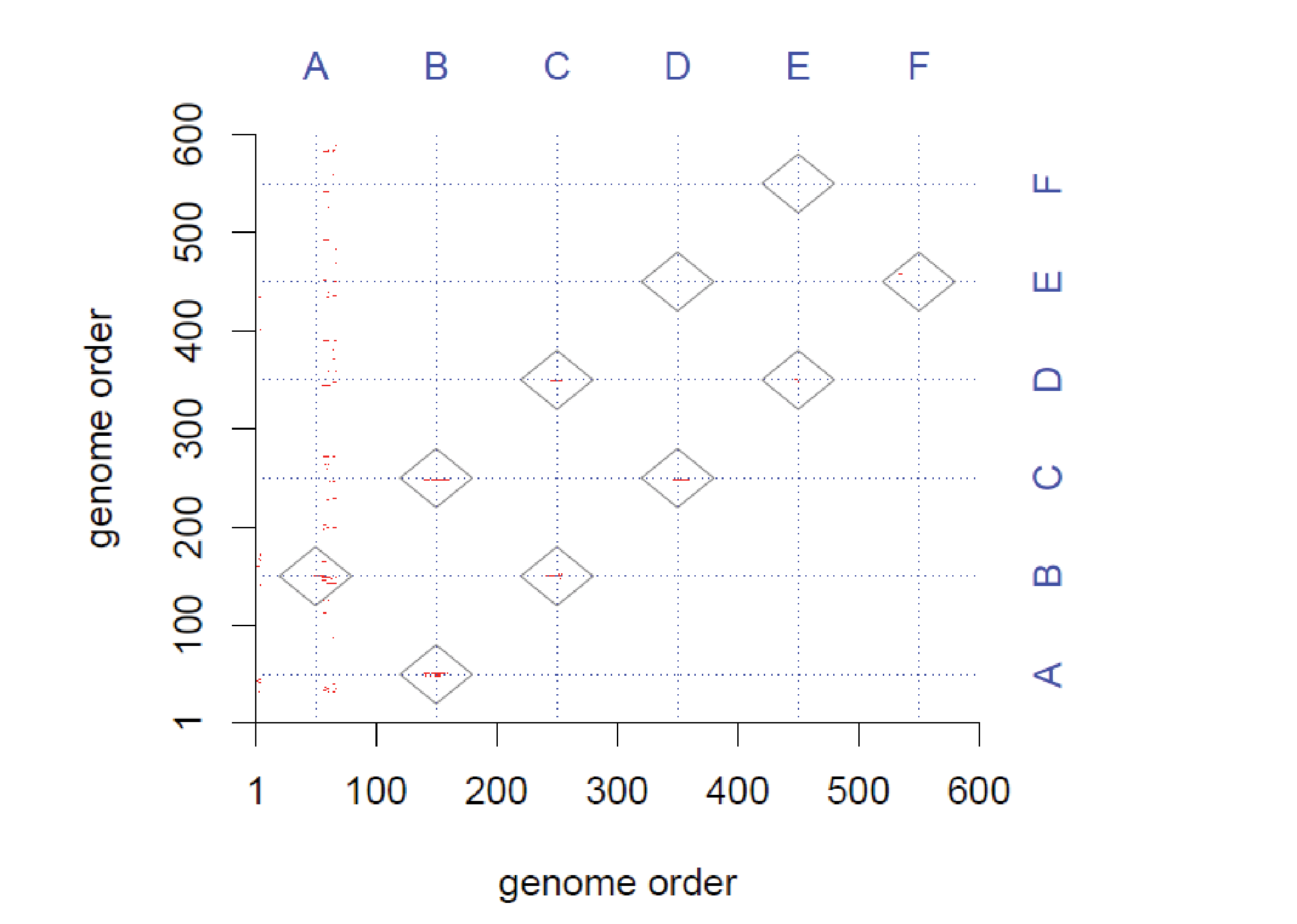}}
\caption{Results of \texttt{LogitNet} and \texttt{SepLogit} on the
example data set shown in Supplemental Figure S-1. Each red dot
represents a non-zero $\beta_{rs}$. Points in the grey diamond are
deemed as correct detections. The dashed blue lines indicate the
locations of aberration \texttt{A-F}.} \label{Fig:ChainBeta}
\end{center}
\vspace{-23pt}
\end{figure}

For the chain model, aberrations \texttt{A}-\texttt{F} occur
sequentially on one oncogenic pathway. The aberration frequencies
and transitional probabilities along the oncogenic pathway are
illustrated in Figure~\ref{Fig:ChainModel}. The true conditionally
dependent pairs in this model are
$$E = \{(s_\mathtt{A}, s_\mathtt{B}), (s_\mathtt{B}, s_\mathtt{C}),
(s_\mathtt{C}, s_\mathtt{D}), (s_\mathtt{D}, s_\mathtt{E}), (s_\mathtt{E},
s_\mathtt{F})\}.$$
Based on this chain model, we generated 50 independent data sets.
The heatmap of one example data matrix $200\times
600$ is shown in Supplemental Figure S-1. We then
apply \texttt{LogitNet} and \texttt{SepLogit} to each simulated data
set for a series of different values of 
$\lambda$. Figure~\ref{Fig:Chain.error} shows the \texttt{FPR} and
\texttt{FNR} of the two methods as a function of $\lambda$.
For both methods, \texttt{FPR} decreases with $\lambda$ while
\texttt{FNR} increases with $\lambda$. Comparing the two methods,
\texttt{LogitNet} clearly outperforms \texttt{SepLogit} in terms of
\texttt{FPR} and \texttt{FNR}. For \texttt{LogitNet}, the average
optimal total error rate (\texttt{FPR}$+$\texttt{FNR}) across the 50
independent data sets is 0.014 (s.d.=0.029); while the average optimal total error
rate for \texttt{SepLogit} is 0.211 (s.d.=0.203). Specifically, taking
the data set shown in the Supplemental Figure S-1 as an example, the
optimal total error rate achieved by \texttt{LogitNet} on this data
set is 0, while the optimal total error achieve by \texttt{SepLogit}
is 0.563 (\texttt{FPR}$=0.563$, \texttt{FNR}$=0$). The corresponding
two coefficient matrices $\hat{\cal{B}}$ are illustrated in
Figure~\ref{Fig:ChainBeta}. As one can see, there is
a large degree of asymmetry in the result of \texttt{SepLogit}: 435
out of the 476 non-zero $\hat\beta_{rs}$ have inconsistent transpose
elements, $\hat\beta_{sr}=0$.
On the contrary, by enforcing symmetry our proposed approach
\texttt{LogitNet} has correctly identified all five true conditionally
dependent pairs in the chain model.  Moreover, the non-zero
$\hat\beta_{rs}$'s plotted by red dots tend to be clustered within
the grey diamonds.  This shows that \texttt{LogitNet} indeed encourages
group selection for highly correlated predictors, and thus is
able to make good use of the  spatial correlation in the data when
inferring the edges.

We also evaluated the two penalty parameter selection approaches: CV
and BIC, for \texttt{LogitNet}.
Table~\ref{table:tuning} summarizes the \texttt{FPR}
and \texttt{FNR} for CV and BIC.  Both approaches performed
reasonably well. The CV criterion tends to select larger models than
the BIC, and thus has more false positives and fewer false
negatives. The average total error rate (\texttt{FPR}$+$\texttt{FNR})
for CV is 0.079, which is slightly smaller than the total error rate
for BIC, 0.084.

\begin{table}\caption{Summary of \texttt{FPR}
and \texttt{FNR} for \texttt{LogitNet} for using CV and BIC to choose optimal
$\lambda$. Each entry is the mean (S.D.) over 50 independent data sets}
\begin{tabular}{c|cc|cc}\hline
 & \multicolumn{2}{|c|}{Chain Model} & \multicolumn{2}{|c}{Tree Model}
 \\\hline
 & \texttt{FPR} & \texttt{FNR} & \texttt{FPR} & \texttt{FNR}\\
 \texttt{CV} &   0.079 (0.049) &  0 (0)   &  0.058 (0.059)   & 0.280 (0.17)\\
 \texttt{BIC} &  0.025 (0.035)  &  0.06 (0.101)   &  0.024 (0.038) & 0.436 (0.197)\\
\hline
\end{tabular}
\label{table:tuning}
\end{table}

\subsection{Simulation II --- Tree Model}
\begin{figure}[h]
\begin{center}
 \subfigure[False positive rate (\texttt{FPR}).]{
\includegraphics[height=14cm, angle=270]{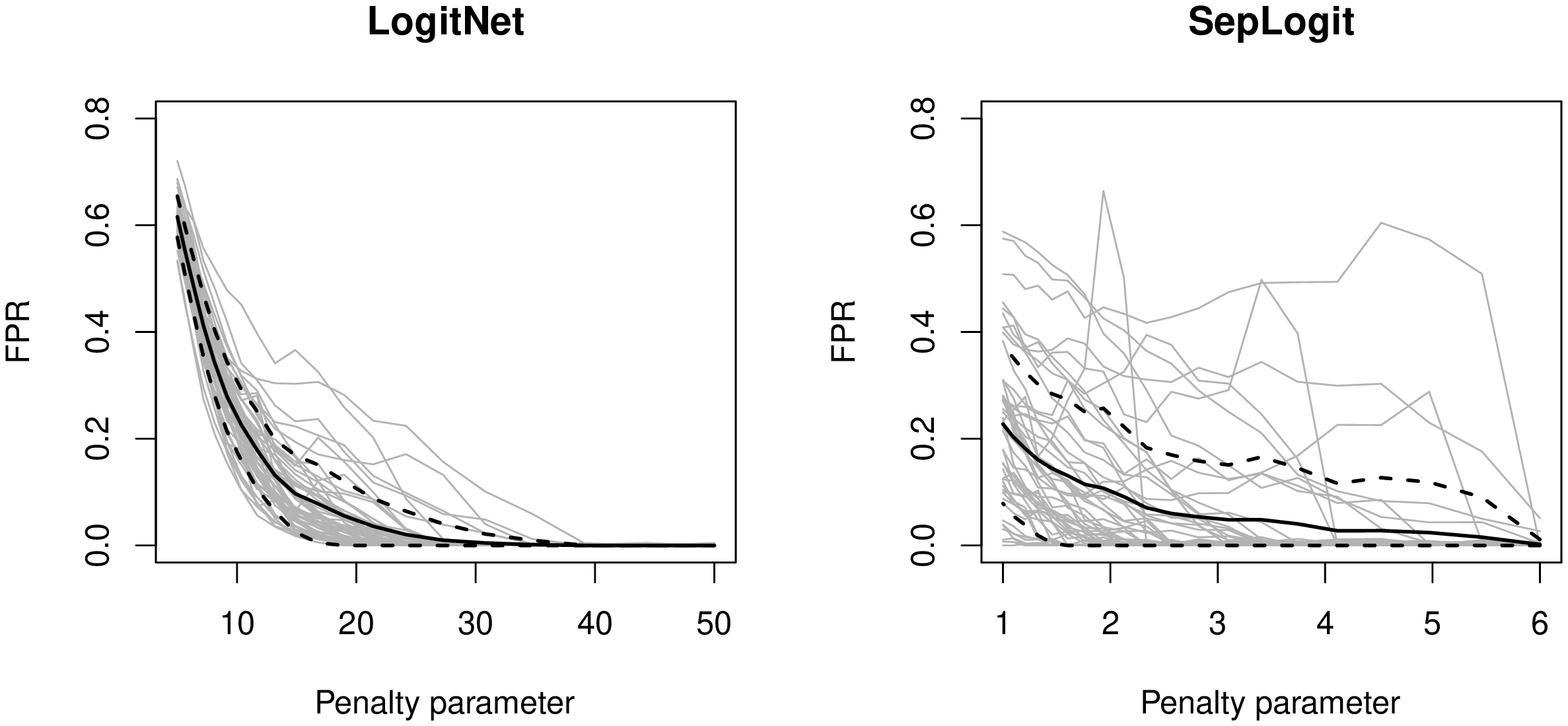}}

\subfigure[False negative rate (\texttt{FNR}).]{
 \includegraphics[height=14cm, angle=270]{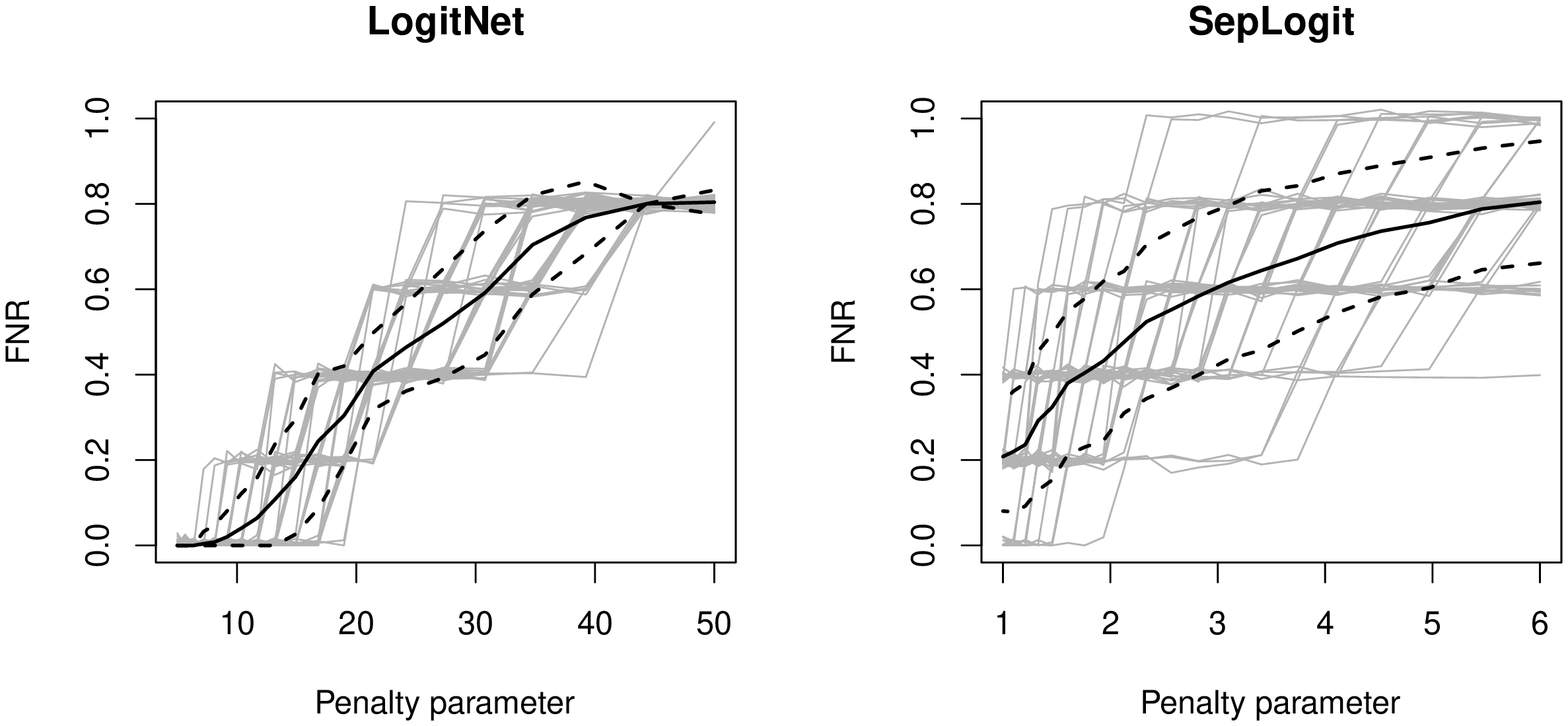}}
 \subfigure[Total error rate (\texttt{FPR}+\texttt{FNR}).]{
 \includegraphics[height=14cm, angle=270]{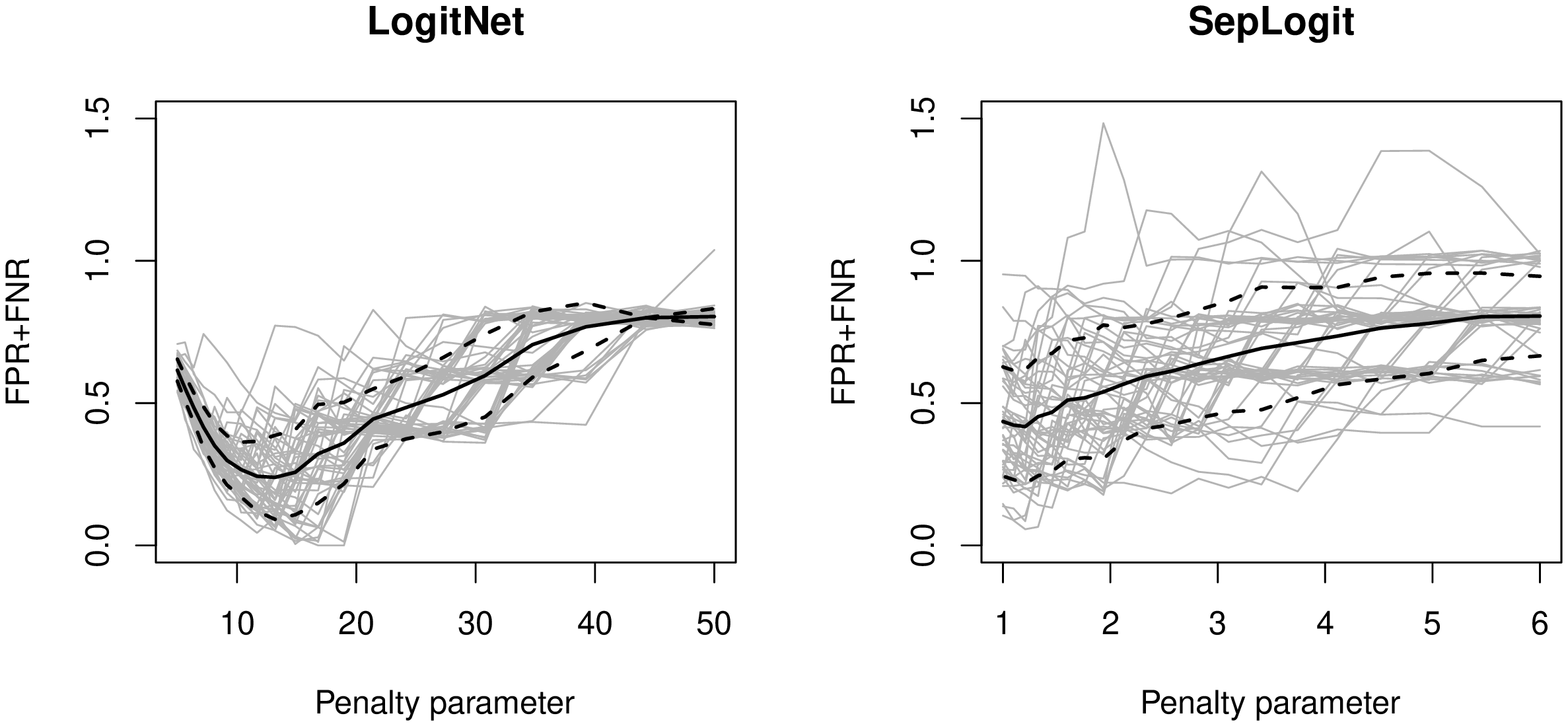}}
 \caption{Results of \texttt{LogitNet} and \texttt{SepLogit} for 
the tree model in Figure~\ref{Fig:TreeModel}. Each
grey line represents one of the 50 independent data
sets. The solid line is the mean curve with the two dashed
lines represent mean $\pm$ one s.d.}
\label{Fig:tree.error}
\end{center}
\end{figure}

For the tree model, we used the empirical mutagenic tree derived
in Beerenwinkel et al.\ (2004) for a HIV data set. The details of the
model are illustrated in Figure~\ref{Fig:TreeModel}. The true
conditionally dependent pairs in this model are
$$ E =\{(s_\mathtt{A}, s_\mathtt{B}), (s_\mathtt{B}, s_\mathtt{E}),
(s_\mathtt{A}, s_\mathtt{C}), (s_\mathtt{C}, s_\mathtt{F}),
(s_\mathtt{A}, s_\mathtt{D})\}.$$ The results of \texttt{LogitNet}
and \texttt{SepLogit} for these data sets are summarized in
Figure~\ref{Fig:tree.error}. Again, \texttt{LogitNet} outperforms
\texttt{SepLogit} in terms of \texttt{FPR} and \texttt{FNR}. The
average optimal total error rate (\texttt{FPR}+\texttt{FNR})
achieved by \texttt{LogitNet} across the 50 independent data sets is
0.163 (s.d.=0.106); while the average optimal total error rate for
\texttt{SepLogit} is 0.331 (s.d.=0.160), twice as large as
\texttt{LogitNet}. We also evaluated CV and BIC for
\texttt{LogitNet}. The 
results are summarized in Table~\ref{table:tuning}. Both CV and BIC
give much higher \texttt{FNR}s under the tree model than under
the chain model. This is not surprising as some transition
probabilities between aberration events along the pathway are smaller
in the tree model than in the chain model.  As in
the chain model, we also observe that BIC gives
smaller \texttt{FPR} and higher \texttt{FNR} than CV, suggesting CV
tends to select larger models and thus yields less
false negatives but with more false positives in detecting edges.


\section{Application to a Breast Cancer Data Set}\label{sec:RealApplication}

In this section, we illustrate our method using a genomic
instability data set from breast cancer samples.  In this data set
the genomic instability is measured by loss of heterozygosity (LOH),
one of the most common alterations in breast cancer. An LOH event at
a marker locus for a tumor is defined as a locus that is homozygous
in the tumor and heterozygous in the constitutive normal DNA.   To
gain a better understanding of LOH in breast cancer, Loo et al.
(2008) conducted a study which used the GeneChip Mapping 10K Assay
(Affymetrix, Santa Clara, CA) to measure LOH events in 166 breast
tumors derived from a population-based sample. The array contains
9706 SNPs, with 9670 having annotated genome locations.
Approximately 25\% of the SNPs are heterozygous in normal DNA, which
means LOH can not be detected in the remaining 75\% of SNPs, i.e.,
the SNPs are non-informative. To minimize the missing rate for
individual SNPs, we binned the SNPs by cytogenetic bands (cytoband).
A total of 765 cytobands are covered by these SNPs. For each sample,
we define the LOH status of a cytoband  to be 1 if at least 2
informative SNPs in this cytoband show LOH and 0 otherwise. We then
remove 164 cytobands which either have missing rates above $20\%$,
or show LOH in less than 5 samples to exclude rare events. The
average LOH rate in the remaining 601 cytobands is $12.3\%$.


Despite our effort to minimize missingness in the data,  $7.5\%$ of
values are still missing in the remaining data. We use the
multiple imputation approach to impute the  missing values based on
the conditional probability of LOH at the target SNP given the available LOH
status at adjacent loci.  If both adjacent loci are missing LOH
status, we will impute the genotype using only the marginal
probability of the target SNP.  See Supplemental Appendix D for
details of the multiple imputation algorithm.  

We then generate 10 imputed data sets. We apply \texttt{LogitNet} on
each of them and use 10-fold CV for penalty parameter selection. The
total number of edges inferred on each imputed data set is
summarized in Table~\ref{table:edgeNumber}. We can see that two
imputation data sets have far more edges detected than the rest of
imputation data sets. This suggests that there is a substantial
variation among imputed data sets and we can not reply on a single
imputed data set.  Thus, we examine the consensus edges across
different imputation runs. There are 3 edges inferred in at least 4
imputed datasets (Table~\ref{table:interaction}). Particularly,
interaction between 11q24.2 and 13q21.33 has been consistently
detected in all of the 10 imputation data sets. Detailed numbers of
LOH frequencies at these two cytobands are shown in Supplementary
Table S-1. Cytoband 11q24.2 harbors the CHEK1 gene, which is an
important gene in the maintenance of genome integrity and a
potential tumor suppressor. DACH1  is located on cytoband 13q21.33 and
has a role in the inhibition of tumor progression and metastasis in
several types of cancer (e.g., Wu et al., 2009).  Both CHEK1
and DACH1 inhibit cell cycle progression through mechanisms involving
the cell cycle inhibitor, CDKN1A.  See Supplemental Figure S-2 for the
pathway showing the interaction between CHEK1 on 11q24.2 and DACH1
on 13q21.33.  

\begin{table}
\caption{Number of edges detected in each imputed data set. }
\begin{tabular}{c|cccccccccc}\hline
 Imputation Index & 1 & 2 & 3 & 4 & 5 & 6 & 7 & 8 & 9 & 10\\
 $\#$ of edges detected & 2  & 2 &  5 & 219 &  3 & 10 &  1 & 114 &  2 &  1\\
\hline
\end{tabular}
\label{table:edgeNumber}
\end{table}

\begin{table}\caption{Annotation for the edges inferred in at least 4
 out of 10 imputed datasets. }
\begin{tabular}{cc} \hline
Cytoband pair &  Frequency of selection \\\hline
11q22.3, 13q33.1   & 6\\
11q24.2, 13q21.33  & 10\\
11q25, 13q14.11 & 4\\
\hline
\end{tabular}
\label{table:interaction}
\end{table}

\section{Final Remarks}
In this paper, we propose the \texttt{LogitNet} method for learning
networks using high dimensional binary data. The work is motivated
by the interest in inferring disease oncogenic pathways from genomic
instability profiles (binary data).  We show that under the
assumption of no cycles for the oncogenic pathways, the dependence
parameters in the joint probability distribution of binary variables
can be estimated by fitting  a set of logistic regression models
with a symmetric coefficient matrix.  For
high-dimension-low-sample-size data, this result is especially
appealing as we can use sparse regression techniques to regularize
the parameter estimation.  We implemented a fast algorithm for
obtaining the \texttt{LogitNet} estimator.  This algorithm enforces
the symmetry of the coefficient matrix and also accounts for the
spatial correlation in the genomic instability profiles by a
weighting scheme.  With extensive simulation studies, we demonstrate
that this method achieves good power in edge detection, and also
performs favorably compared to an existing method.


In \texttt{LogitNet}, the weighting scheme together with the
enforcement of symmetry encourage a group selection effect,
\textit{i.e.}, highly spatially correlated variables tend to be in and
out of the model simultaneously.  It is conceivable that this
group selection effect may be further enhanced by replacing the
\texttt{lasso} penalty with the \texttt{elastic net} penalty proposed
by Zou and Haste (2005) as $ \lambda_1\sum_{1\leq r<s\leq p}|\beta_{rs}|
+\lambda_2\sum_{1\leq r<s\leq p}\beta_{rs}^2$. The square $\ell_2$ norm
penalty may facilitate group selection within each regularized
logistic regression. More investigation along this line
is warranted.

R package \texttt{LogitNet} is available from the authors upon
request.  It will also be made available through CRAN shortly.

\begin{center} {\sc Acknowledgments}
\end{center}

The authors are grateful to Drs. Peggy Porter and Lenora Loo for
providing the genomic instability data set to us, which has motivated
this methods development work.  The authors are in part supported by
grants from the National Institute of Health,  R01GM082802 (Wang),
R01AG14358 (Chao and Hsu), and P01CA53996 (Hsu).




\begin{thebibliography}{99}

\bibitem{Beerenwinkel05} Beerenwinkel, N., Rahnenführer, J., D\"{a}umer, M.,
Hoffmann, D., Kaiser, R., Selbig, J. and Lengauer, T. (2005). Learning
Multiple Evolutionary Pathways from Cross-Sectional Data. \textit{Journal 
of Computational  Biology} \textbf{12}, 584--598.


\bibitem{cox72} Cox, D.R. (1972). The analysis of multivariate binary
  data.  \textit{Applied Statistics} \textbf{21},
  113--120.

\bibitem{dawid93} Dawid, A.P. and Lauritzen, S.L. (1993).  Hyper-Markov laws
  in the statisticla analysis of decomposable graphical models. \textit{Annals of Statistics}  \textbf{21}, 1272--1317.

\bibitem{drton04} Drton, M. and  Perlman, M.D. (2004). Model selection for
  Gaussian concentration graphs. \textit{Biometrika} \textbf{91}, 591--602.


\bibitem{Edward00} Edward, D. (2000). Introduction to Graphical Modelling (2nd ed.),
\textit{New York: Springer}.

\bibitem{Efron04} Efron, B., Hastie, T., Johnstone, I., and Tibshirani, R.
(2004). Least Angle Regression. \textit{Annals of Statistics} \textbf{32},
407--499.

\bibitem{Friedman07a} Friedman, J., Hastie, T., Hofling, H., and Tibshirani, R. (2007a).
Pathwise Coordinate Optimization. \textit{Annals of Applied
Statistics}, \textbf{1}, 302--332.

\bibitem{Friedman07b} Friedman, J., Hastie, T. and Tibshirani, R. (2007b).
Sparse inverse covariance estimation with the graphical lasso.
\textit{Biostatistics} \textbf{9}, 432--441.

\bibitem{Friedman08} Friedman, J., Hastie, T. and Tibshirani, R. (2009). Regularization Paths for Generalized Linear Models via Coordinate Descent.
\textit{Technical report}: http://www- stat.stanford.edu/
jhf/ftp/glmnet.pdf.

\bibitem{Fu98} Fu, W. (1998). Penalized Regressions: the Bridge vs the
Lasso. \textit{Journal of Computational and Graphical Statistics}
 \textbf{7}, 397--416.

\bibitem{genkin07} Genkin, A., Lewis, D.D., Madigan,
  D. (2007). Large-scale Bayesian logistic regression for text
  categorization.  \textit{Technometrics} \textbf{49}, 291--304.

\bibitem{joe96} Joe, H. and Liu, Y. (1996).  A model for a
multivariate binary response with covariates based on compatible
conditionally specified logistic regression.  \textit{Statistics \&
Probability Letters} \textbf{31},   113--120.

\bibitem{Klein85} Klein, G. and Klein, E. (1985). Evolution of tumors and the
  impace of molecular oncology. \textit{Nature} \textbf{315}, 190--195.


\bibitem[\protect\citeauthoryear{Lauritzen}{Lauritzen}{1996}]{Lauritzen1996}
Lauritzen, S.L. (1996). \textit{Graphical Models} Clarendon Press, Oxford, United Kingdom.

\bibitem{Li06} Li, H. and Gui, J. (2006). Gradient Directed Regularization for
Sparse Gaussian Concentration Graphs, with Applications to Inference
of Genetic Networks. \textit{Biostatitics}, \textbf{7}, 302--317.


\bibitem{loo08} Loo, L., Ton, C., Wang, Y.W., Grove, D.I., Bouzek, H., Vartanian, N., Lin, M.G.,
  Yuan, X., Lawton, T.L., Daling, J.R., Malone, K.E., Li, C.I., Hsu, L.,
  Porter, P. (2008).  Differential patterns of allelic loss in estrogen
  receptor-positive infiltrating lobular and ductal breast cancer.
  \textit{Genes, Chromosomes and Cancer} \textbf{47}, 1049--66.

\bibitem{madigan95} Madigan, D. and York, J. (1995). Bayesian graphical
  models for discrete data. \textit{International Statistical Review}
  \textbf{63}, 215--232.

\bibitem{meinshausen06} Meinshausen, N. and Buhlmann, P. (2006).  High
  dimensional graphs and variable selection with the Lasso.  \textit{Annals of
  Statistics} \textbf{34},  1436--1462.


\bibitem{Newton98} Newton, M.A., Gould, M.N., Reznikoff, C.A. and Haag, J.D.
  (1998).  On the   statistical analysis of allelic-loss
  data. \textit{ Statistics in  Medicine} \textbf{17}, 1425--1445.

\bibitem{Peng09} Peng, J., Wang, P., Zhou, N. and Zhu, J. (2009a). Partial Correlation
Estimation by Joint Sparse Regression Model. \textit{Journal of the American Statistical Association} \textbf{104}, 735--746.

\bibitem{Peng09b} Peng, J., Zhu, J., Bergamaschi, A., Han, W., Noh, D.Y., Pollack, J.R.,
Wang, P. (2009b). Regularized Multivariate Regression for Identifying
Master Predictors with Application to Integrative Genomics Study of
Breast Cancer. \textit{Technique Report}
http://arxiv.org/abs/0812.3671.

\bibitem{prentice91} Prentice, R.L. and Zhao, L.P. (1991). Estimating
  equations for parameters in means and   covariances of multivariate
  discrete and continuous responses.     \textit{Biometrics}
  \textbf{47}, 825--839.

\bibitem{Ravikumar09} Ravikumar, P., Wainwright, M. and Lafferty, J.
(2009). High-dimensional Ising model selection using
$l_1$-regularized logistic regression. \textit{Annals of Statistics}
to appear.

\bibitem{Rothman08} Rothman, A.J., Bickel, P.J., Levina, E. and Zhu, J.
  (2008). Sparse permu- tation invariant covariance
  estimation. \textit{Electronic Journal of Statistics} \textbf{2}, 494--515.

\bibitem{Shafer07} Schafer, J. and Strimmer, K. (2005). A Shrinkage Approach to
Large-Scale Co- variance Matrix Estimation and Implications for
Functional Genomics. \textit{Statistical Applications in Genetics
and Molecular Biology}, \textbf{4}, Article 32.

\bibitem{tibshirani96} Tibshirani, R. (1996).  Regression shrinkage and
  selection via the lasso.  \textit{Journal of the Royal Statististical Society Series B} \textbf{58}, 267--88.


\bibitem{Whittaker90} Whittaker, J. (1990). \textit{Graphical Models in Applied Mathematical
Multivariate Statistics.} {Wiley}.

\bibitem{Wu09} Wu, K.,  Katiyar, S., Witkiewicz, A., Li, A., McCue, P., Song, L.,
  Tian, L., Jin, M., Pestell, R.G. (2009) The Cell Fate Determination Factor
  Dachshund Inhibits Androgen Receptor Signaling and Prostate Cancer
  Cellular Growth.    \textit{Cancer Research} \textbf{69}, 3347--3355.


\bibitem{Yuan07} Yuan, M. and Lin, Y. (2007). Model Selection and Estimation in the
Gaussian Graphical Model. \textit{Biometrika} \textbf{94}, 19--35.

\bibitem{zhang01} Zhang, T. and Oles, F. (2001).  Text categorization
  based on regularized linear classifiers.  \textit{Information Retrieval} \textbf{4},
  5--31.

\bibitem{zhao90} Zhao, L.P. and Prentice, R.L. (1990). Correlated
  binary regression using a quadratic exponential model.
  \textit{Biometrika} \textbf{77}, 642--648.

\bibitem{zou04} Zou, H., Hastie, T. (2005). Regularization and variable
  selection via the elastic net. \textit{Journal of the Royal Statistical Society, Series B} \textbf{67},
  301--320.

\bibitem{zou07} Zou, H., Hastie, T., Tibshirani, R. (2007). On   the degrees of
 freedom of the lasso. \textit{Annals of Statistics} \textit{35} 2173--2192.
\end{thebibliography}
\end{document}